\documentclass[aps,prd,preprint,superscriptaddress,tightenlines,nofootinbib,showpacs]{revtex4}
\usepackage{epsfig}
\usepackage{afterpage}
\usepackage{graphicx}% Include figure files
\usepackage{dcolumn}% Align table columns on decimal point
\usepackage{bm}% bold math

\newcommand{\etal}{{\it et al.}}

\begin{document}

%\begin{flushright}
%Paper Draft\\
%S.~Khalil\\
%S.~Stone~\\
%L.~Zhang~\\
% \today
%\end{flushright}

\preprint{\tighten\vbox{\hbox{\hfil CLNS 08/2029}
                        \hbox{\hfil CLEO 08-12}}}

\title{Precision Measurement of {\boldmath ${\cal{B}}(D^+\to\mu^+\nu)$} and the
Pseudoscalar Decay Constant {\boldmath $f_{D^+}$}}

%Improved Measurement of B(D+ -> mu+nu and the Pseudoscalar Decay Constant fD+
%-------- INSERT HERE ------------

%\author{S. Khalil}
%\author{L. Zhang}

%\author{S.~Stone}

%\affiliation{Syracuse University, Syracuse, New York 13244}
\author{B.~I.~Eisenstein}
\author{I.~Karliner}
\author{S.~Mehrabyan}
\author{N.~Lowrey}
\author{M.~Selen}
\author{E.~J.~White}
\author{J.~Wiss}
\affiliation{University of Illinois, Urbana-Champaign, Illinois 61801, USA}
\author{R.~E.~Mitchell}
\author{M.~R.~Shepherd}
\affiliation{Indiana University, Bloomington, Indiana 47405, USA }
\author{D.~Besson}
\affiliation{University of Kansas, Lawrence, Kansas 66045, USA}
% ...
\collaboration{CLEO Collaboration}
\noaffiliation

\date{June 4, 2008}

\begin{abstract}
We measure the branching ratio of the purely leptonic decay of the
$D^+$ meson with unprecedented precision as
${\cal{B}}(D^+\to\mu^+\nu)=(3.82\pm 0.32\pm 0.09)\times 10^{-4}$,
using 818 pb$^{-1}$ of data taken on the $\psi(3770)$ resonance with
the CLEO-c detector at the CESR collider. We use this determination
to derive a value for the pseudoscalar decay constant $f_{D^+}$,
combining with measurements of the $D^+$ lifetime and assuming
$|V_{cd}|=|V_{us}|$. We find $f_{D^+}=(205.8\pm 8.5\pm 2.5)~{\rm
MeV}$. The decay rate asymmetry
$\frac{\Gamma(D^+\to\mu^+\nu)-\Gamma(D^-\to\mu^-\bar{\nu})}
{\Gamma(D^+\to\mu^+\nu)+\Gamma(D^-\to\mu^-\bar{\nu})} =
0.08\pm0.08$, consistent with no CP violation.
 We also set 90\% confidence level upper limits on
 ${\cal{B}}(D^+\to \tau^+\nu)<1.2\times 10^{-3}$ and
${\cal{B}}(D^+\to e^+\nu)<8.8\times 10^{-6}$.
\end{abstract}


\maketitle
%\tighten

%\newpage
\section{Introduction}
\begin{figure}[htb]
%\vskip 0.00cm \centerline{ \epsfxsize=6.0in
\centerline{ \epsfxsize=3.0in \epsffile{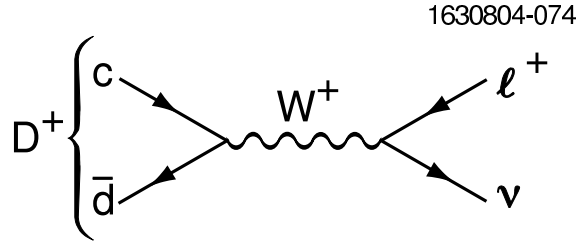} } \caption{The
decay diagram for $D^+\to \ell^+\nu$.} \label{Dptomunu}
\end{figure}

Purely leptonic decays of heavy mesons involve both weak and strong
interactions. The weak part is easy to describe as the annihilation
of the quark antiquark pair via the Standard Model $W^+$ boson; the
Feynman diagram for $D^+\to \ell^+\nu$ is shown in
Fig.~\ref{Dptomunu}.
The strong interactions arise due to gluon exchanges between the
charm quark and the light quark. These are parameterized in terms of
the ``decay constant" for the $D^+$ meson $f_{D^+}$. The
decay rate is given by \cite{Formula1}
\begin{equation}
\Gamma(D^+\to \ell^+\nu) = {{G_F^2}\over
8\pi}f_{D^+}^2m_{\ell}^2M_{D^+} \left(1-{m_{\ell}^2\over
M_{D^+}^2}\right)^2 \left|V_{cd}\right|^2~~~, \label{eq:equ_rate}
\end{equation}
where $G_F$ is the Fermi coupling constant, $M_{D^+}$ is the $D^+$
mass, $m_{\ell}$ is the mass of the final state lepton, and $V_{cd}$ is
a Cabibbo-Kobayashi-Maskawa (CKM) matrix element \cite{PDG},
whose magnitude is set equal to 0.2256, the
value of $V_{us}$ \cite{Antonelli}. Thus, within the context of the
Standard Model (SM), measurement of this purely leptonic decay
provides a means of determining $f_{D^+}$, and similarly measuring
the purely leptonic decay of the $D_s^+$ meson allows us to
determine $f_{D_s}$.

Meson decay constants in the $B$ system are used to translate
measurements of $B\bar{B}$ mixing to CKM matrix elements. Currently,
it is not possible to determine $f_B$ accurately from leptonic $B$
decays, so theoretical calculations of $f_B$ must be used. Since the
$B_s$ meson does not have $\mu\nu$ decays, it will never be possible
to determine $f_{B_s}$ experimentally, so again theory must be
relied upon. If calculations disagree on $D$ mesons, they may be
questionable on $B$ mesons. If, on the other hand new physics is
present, it is imperative to understand how it affects SM-based
predictions of the $B$ decay constants.

These decay constants can be calculated in theories of QCD. A recent
calculation by Follana~\etal ~\cite{Lat:Follana} using an unquenched
lattice technique predicts $f_{D^+}=(207\pm 4)$ MeV and
$f_{D_s}=(241\pm 3)$ MeV. The latter result differs by more than
three standard deviations from the average of CLEO and Belle
measurements \cite{Rosner-Stone}.

Dobrescu and Kronfeld point out that this discrepancy can be caused
by the presence of non-SM objects participating virtually in the
decay \cite{Dobrescu-Kron}. They give three possibilities: (1) a new
boson of charge +1 interfering with the SM $W^+$ annihilation, (2) a
charge $+2/3$ leptoquark, and (3) a charge minus $1/3$ leptoquark. The charge +1
boson could either be a $W'^+$ or a charged Higgs. They propose a
specific two Higgs doublet model where one doublet gives the $c$,
$u$  and leptons mass, but not the $d,~s,~b,$ or $t$, and has a
vacuum expectation value of about 2 GeV. Such a model predicts that
the ratio of widths
$\Gamma(D_s^+\to\tau^+\nu)$/$\Gamma(D_s^+\to\mu^+\nu)$ is the same
as the Standard Model expectation, which
is in agreement with the CLEO measurements.

The previous CLEO determination of $f_{D^+}=(222.6\pm
16.7^{+2.3}_{-3.4})$ MeV is consistent with the Follana \etal
~calculation at the one standard deviation level, but the
experimental error was too large to provide a precision test. Here we
provide a measurement based on a three times larger data sample and
a $\approx$15\% larger efficiency based on improved analysis
techniques.

One other fully unquenched lattice calculation exists in the
literature \cite{Lat:Milc}, although it has significantly larger
errors than Follana \etal~\cite{Lat:Follana}. Quenched calculations
have also been performed \cite{QCDSF,Lat:Taiwan,Lat:UKQCD,Lat:Damir}, and other
methods have been used
\cite{Bordes,Equations,Field,Chiral,Sumrules,Quarkmodel,Isospin}. The
various theoretical predictions of $f_{D^+}$ range from 190 MeV to
350 MeV. Because of helicity suppression, the electron mode $D^+ \to
e^+\nu$ has a very small rate in the Standard Model \cite{Akeroyd}.
The expected relative widths are $2.65:1:2.3\times 10^{-5}$ for the
$\tau^+ \nu$, $\mu^+ \nu$, and $e^+ \nu$ final states, respectively.
Unfortunately the mode with the largest branching fraction,
 $\tau^+\nu$, has at least
 two neutrinos in the final state and is difficult to detect in $D^+$ decay.

The CLEO-c detector is equipped to measure the momenta and directions
of charged particles, identify charged hadrons, detect photons, and
determine their directions and energies  with good precision. It has
been described in more detail previously
\cite{CLEODR,RICH,CLEOD,CLEODptomunu}.
%  \cite{CLEODR}, \cite{RICH}, \cite{CLEOD}, \cite{CLEODptomunu}.

\section{Data Sample and Signal Selection}

%\subsection{Data Sample}
In this study we use 818 pb$^{-1}$ of CLEO-c data collected from
$e^+e^-$ collisions at the $\psi(3770)$ resonance.
This work contains our previous sample as a subset and
supersedes our initial efforts \cite{CLEODptomunu}.
 At this energy, the events consist mostly of pure
$D^+D^-$, $D^0\overline{D}^0$, three-flavor continuum, with
small amounts of other final states such as
$\gamma\psi(2S)$ and $\tau^+\tau^-$.

We examine all the recorded hadronic events and retain those containing at
least one charged $D$ candidate in the modes listed in
Table~\ref{tab:Dreconnew}. We use this sample to look for cases
where we have only a single muon candidate whose four-momentum is
consistent with a two-body $D$ decay into a muon and a neutrino and
no other charged tracks or excess neutral energy are present. Track
selection, particle identification, $\pi^0$, $K_S$ and muon
selection criteria are identical to those described in Reference
\cite{CLEODptomunu}, with one important exception. The angular
acceptance of the muon has been widened to cover 90\% of the
solid angle rather than 81\%. Muons deposit less than 300 MeV of
energy in the calorimeter 98.8\% of the time, while hadrons often
interact and deposit significantly more energy.  Thus, we define two cases in
this paper, where case (i) refers to muon candidate tracks that
deposit $<$300 MeV and case (ii) is for candidates
depositing $>$ 300 MeV, as was done previously for both our
$D^+\to\tau^+\nu$ and $D_s^+\to\mu^+\nu$ analyses
\cite{Dptotaunu,Dstomunu}. Briefly, we determine the efficiency on
muons from $e^+e^-\to\mu^+\mu^-$ events and compare with our Monte Carlo projection.
The excellent agreement allows us to use the Monte Carlo efficiency for the
lower energy muons observed in this analysis. Pion's deposit $<$300 MeV 55\%
of the time as determined
from a relatively pure sample of $D^0\to K^-\pi^+$ events, and their charge-conjugates.

\section{Reconstruction of Charged {\boldmath $D$} Tagging Modes}
Tagging modes are fully reconstructed by first evaluating the
difference in the energy, $\Delta E$, of the decay products with the
beam energy.  We require the absolute value of this difference to
contain 98.8\% of the signal events, {\it i.e.} to be within $\approx$2.5
times the root mean square (rms) width of the peak value. The rms widths vary from
$\approx$7 MeV in the $K^+K^-\pi^-$ mode to $\approx$14 MeV in the
$K^+\pi^-\pi^-\pi^0$ mode. For the selected events we then view the
reconstructed $D^-$ beam-constrained mass defined as
\begin{equation}
m_{\rm BC}=\sqrt{E_{\rm beam}^2-(\sum_i{\bf p}_{i})^2},
\end{equation}
where $i$ runs over all the final state particles of the tag. Since
the CESR beams have a crossing angle, we work in the center-of-mass
frame. The beam-constrained mass has better resolution than merely
calculating the invariant mass of the decay products since the beam
has a small energy spread. Besides using $D^-$ tags and searching
for $D^+\to\mu^+\nu$, we also use the charge-conjugate $D^+$ tags
and search for $D^-\to \mu^-\overline{\nu}$; in the rest of this
paper we will not usually mention the charge-conjugate modes
explicitly, but they are always used.

\begin{figure}[htbp]
%\vskip 0.00cm \centerline{ \epsfxsize=6.0in
\centerline{ \epsfxsize=6.0in \epsffile{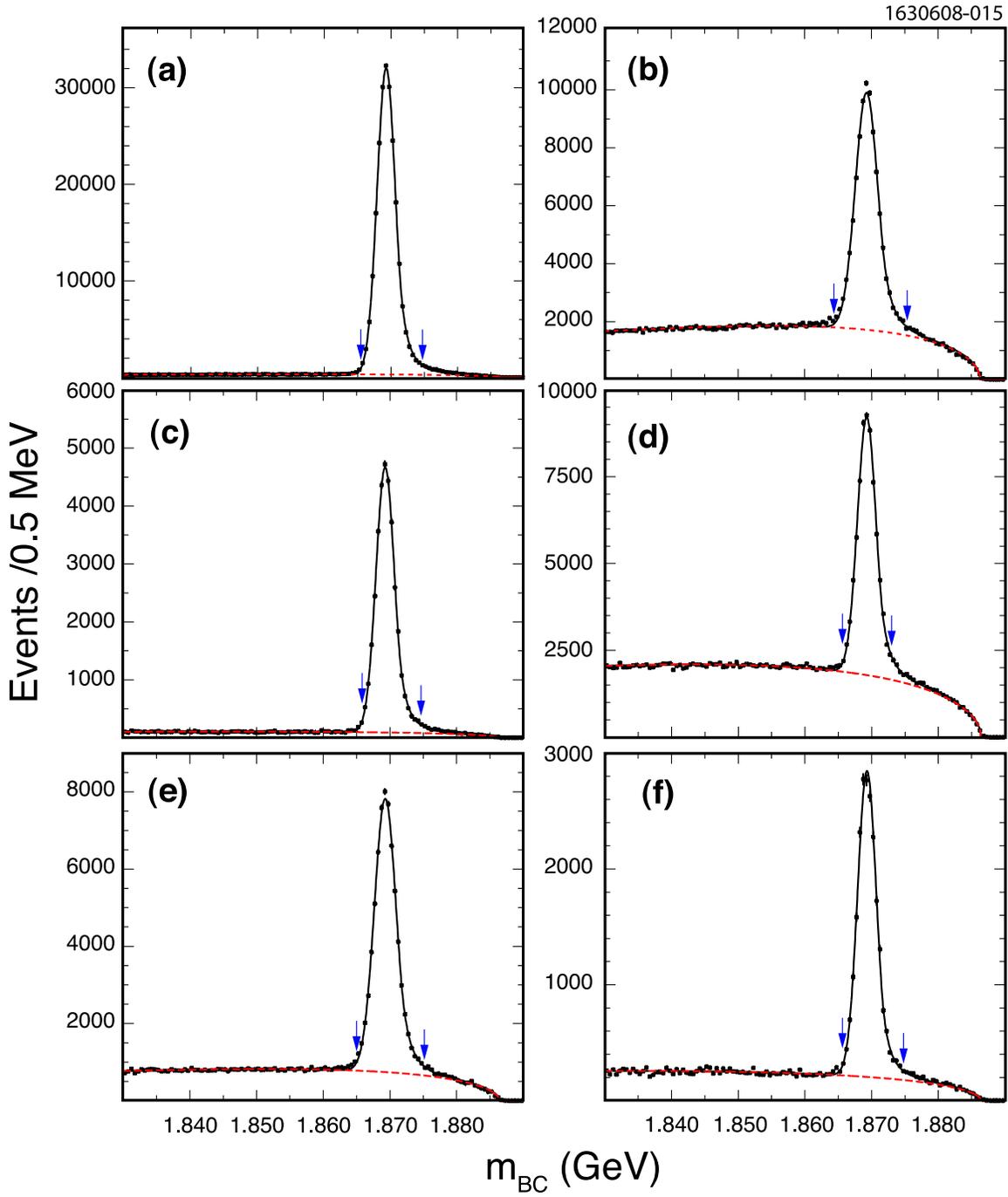} }
\caption{ Beam-constrained mass distributions for different fully
reconstructed $D^-$ decay candidates in the final states: (a) $K^+
\pi^- \pi^-$, (b) $K^+ \pi^- \pi^- \pi^0$, (c) $K_S\pi^-$, (d) $K_S
\pi^-\pi^-\pi^+$, (e) $K_S\pi^- \pi^0$, and (f) $K^+K^-\pi^-$. The
solid curves show the sum of signal and background functions. The
dashed curves indicate the background fits. The region between the
arrows contains the events selected for use in further analysis.}
\label{Dreconnew}
\end{figure}

The $m_{\rm BC}$ distributions for all $D^-$ tagging modes
considered in this data sample are shown in Fig.~\ref{Dreconnew}. To
determine the event numbers we first fit the  $m_{\rm BC}$
distributions to a signal function plus a background shape. Then we
use the signal shape to define the lower and upper limits in $m_{\rm
BC}$, and count the number events above the background function
within the limits.

For the background we fit with a shape function analogous to one
first used by the ARGUS collaboration \cite{ARGUS} which has
approximately the correct threshold behavior at large $m_{\rm BC}$.
This function is
\begin{equation}
f_{\rm background}(m_{\rm BC}) =a(m_{\rm BC}+b)\sqrt{1 - \left({{m_{\rm BC} + b}
\over c} \right)^2} {\rm exp}\left[{ d\left(1 -\left[ {{m_{\rm BC} + b} \over c
}\right]^2\right)}\right],
\end{equation}
where $a$ is the overall normalization and $b$, $c$, and $d$ are
parameters that govern the shape. To fix the shape parameters in
each mode, we fit this function to data selected by using $\Delta E$
sidebands defined as $5\sigma<|\Delta E|< 7.5\sigma$, where $\sigma$
is the rms width of the $\Delta E$ distribution.

For the signal we use a lineshape similar to that used for
extracting photon signals from electromagnetic calorimeters, because
of the tail towards high mass caused by initial state radiation
\cite{CBL}. The functional form is
\[f_{\rm signal}(m_{\rm BC})= \left( \begin{array}{l}
  A\cdot{\rm exp}\left[-{1\over 2}\left({{m_{\rm BC}-m_D}\over \sigma_{m_{\rm BC}}}
\right)^2\right]~~~~{{\rm for}~m_{\rm BC}<m_D-\alpha\cdot\sigma_{m_{\rm BC}}}\\
 A\cdot{{\left({n\over \alpha}\right)^n e^{-{1\over 2}\alpha^2}
\over \left({{m_{\rm BC}-m_D}\over \sigma_{m_{\rm BC}}}+{n\over \alpha}-\alpha\right)^n}}
~~~~~~~~~~~{{\rm for}~m_{\rm BC}>m_D-\alpha\cdot\sigma_{m_{\rm BC}}}\\
\end{array}\right.\]
\begin{equation}
\end{equation}
${\rm Here}~A^{-1}\equiv \sigma_{m_{\rm BC}}\cdot
\left[{n\over \alpha}\cdot {1\over {n-1}}e^{-{1\over 2}\alpha^2}
+\sqrt{\pi\over 2}\left(1+{\rm erf}\left({\alpha\over\sqrt{2}}\right)
\right)\right]$, $m_{\rm BC}$ is the measured mass, $m_D$ is the ``true'' (or most likely) mass
$\sigma_{m_{\rm BC}}$ is the mass resolution, and $\alpha$ and $n$ are shape parameters.

Table~\ref{tab:Dreconnew} lists the modes along with the numbers of
signal events and background events within the signal region defined
as containing 98.8\% of the signal events with $m_{\rm BC}$ below
the peak and 95.5\% of the signal events above the peak.

\begin{table}[htb]
\begin{center}
\caption{Tagging modes and numbers of signal and background events
determined from the fits shown in Fig.~\ref{Dreconnew}.
\label{tab:Dreconnew}}
\begin{tabular}{lcr}\hline\hline
    Mode  &  Signal           &  Background~~ \\ \hline
$K^+\pi^-\pi^- $ & $224,778 \pm 497$   & $5,957$\\
$K^+\pi^-\pi^- \pi^0$ & $71,605\pm 359$  & $37,119$\\
$K_S\pi^-$ &   $32,696\pm 189$& $1,576$\\
$K_S\pi^-\pi^-\pi^+ $ &  $52,554 \pm 315$ & $26,352$\\
$K_S\pi^-\pi^0 $ &  $59,298\pm 289$ & $14,837$\\
$K^+K^-\pi^-$ & 19,124$\pm$159 &3,631 \\\hline
Sum &  $ 460,055\pm 787$ & $89,472$\\
\hline\hline
\end{tabular}
\end{center}
\end{table}

We retain the events within the mass cuts illustrated in
Fig.~\ref{Dreconnew} for further analysis. This sample includes
460,055$\pm$787$\pm$2,760 signal events, where the last error
is systematic.  Because of their low multiplicity, it is easier
to find tags in simple $\mu^+\nu$ events than in typical $D^+D^-$
events. Therefore, to calculate the branching fraction we increase
the number of tags by (1.54$\pm$0.36)\%, as determined by Monte
Carlo simulation. The systematic error on
the signal number includes this uncertainty added in quadrature with the change
given by varying the background function.

\section{{\boldmath $D^+\to \mu^+\nu$} Selection Criteria}
\label{sec:muonsel}
%\subsection{Neutrino Reconstruction}
Using our sample of
$D^-$ event candidates we search for events with a single
additional charged track presumed to be a $\mu^+$. Then we infer
the existence of the neutrino by requiring a measured value
of the missing mass squared (MM$^2$) near
zero (the neutrino mass), where
\begin{equation}
{\rm
MM}^2=\left(E_{\rm beam}-E_{\mu^+}\right)^2-\left(-{\bf p}_{D^-}
-{\bf p}_{\mu^+}\right)^2, \label{eq:MMsq}
\end{equation}
here ${\bf p}_{D^-}$ is the three-momentum of the fully
reconstructed $D^-$, and $E_{\mu^+}$(${\bf p}_{\mu^+}$) is the
energy (momentum) of the candidate $\mu^+$.

To restrict the sample to candidate $\mu^+ \nu$ events resulting
from the other $D$, we exclude events with extra neutral energy, or
more than one additional track with opposite charge to the tagged
$D$, which we take to be the muon candidate. We allow such extra tracks if
their distance of closest approach from the beam collision point is
outside a region more than 5 cm along the beam or more than 5 mm
perpendicular to the beam; we do not wish to veto these tracks as
they are usually due to interactions of the tracks from the tagging
$D^-$ in the calorimeter. We reject events with extra fully
reconstructed $K_S \to\pi^+\pi^-$ candidates. We also veto events
having a maximum neutral energy cluster of more than 250 MeV. This criterion is highly effective in
reducing backgrounds especially from $D^+\to \pi^+\pi^0$ decays.
We consider only those
showers that do not match a charged
 track within a connected region.  A connected region is a group of adjacent crystals
 with finite energy depositions. This reduces the probability of a false
 veto due to hadronic
 shower fragments that would otherwise show up as unmatched
 showers.

Sometimes the decay products of the
tagging $D^-$ interact in the detector material, mostly the EM
calorimeter, and spray tracks and neutral energy back into the rest
of the detector. We evaluate the size of these contributions to the
inefficiency caused by imposing the 250~MeV extra neutral energy requirement by
using fully reconstructed $D^+D^-$ events. We start with events
where the $D^+\to K^-\pi^+\pi^+$ and the $D^-\to K^+\pi^-\pi^-$. We
then look for extra photons with energies $>$250 MeV. This measures
the square of the efficiency for the case of $K^-\pi^+\pi^+$ tags, our
largest mode.
We then measure
the inefficiency for each tag mode by looking for fully
reconstructed events where one $D$ decays into
$K^{\mp}\pi^{\pm}\pi^{\pm}$ and the other into one of the other tag
modes. The weighted average over all our tag modes gives an
efficiency for our extra energy veto of (95.9$\pm$0.2$\pm$0.4)\%. The details are given in
Appendix A.

We define $\theta$ as the angle with respect to the positron beam
direction. The muon candidate direction is required to have
$|\cos\theta|<0.90$, and deposit less than 300 MeV of energy in the
calorimeter, characteristic of a minimum ionizing particle.

\begin{figure}[htbp]
%\vskip 0.00cm \centerline{ \epsfxsize=3.0in
\centerline{ \epsfxsize=3.0in \epsffile{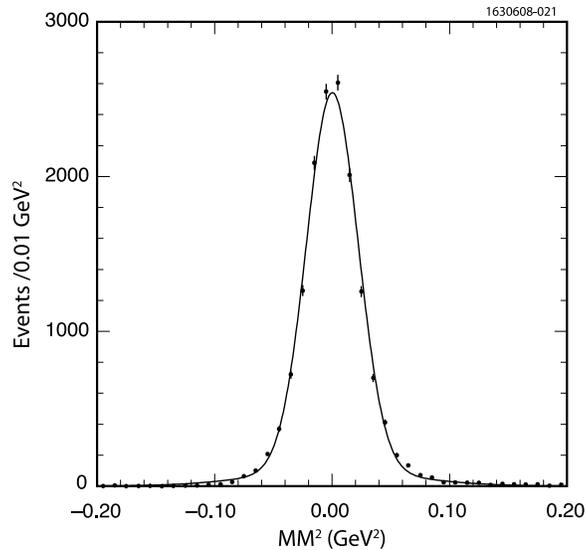}
} \caption{Monte Carlo simulation of the MM$^2$ distributions for
$D^+\to \mu^+ \nu$ events opposite the proper mixture of tag final
states. The fit is to two Gaussian distributions centered at zero
where the second Gaussian constitutes around 30\% of the area.}
\label{mc-mm2}
\end{figure}

The MM$^2$ from Monte Carlo simulation is shown  in
Fig.~\ref{mc-mm2} for the proper mix of tag modes. The signal is fit
to a sum of two Gaussian distributions with the wider Gaussian
having about 30\% of the area independent of tagging mode. The average
resolution ($\sigma$) is defined as
\begin{equation}
\sigma = f_1\sigma_1+(1-f_1)\sigma_2,
\end{equation}
where $\sigma_1$ and $\sigma_2$ are the individual widths of the two
Gaussians and $f_1$ is the fractional area of the first Gaussian.
The resolution of 0.0266$\pm$0.0006 GeV$^2$ is consistent among all
the tagging decay modes when restricting the fit range to
$-0.2<$MM$^2<$0.2 GeV$^2$. In a narrower range, $-0.1<$MM$^2<0.1$
GeV$^2$, the resolution is $\sigma=0.0248\pm 0.0006$ GeV$^2$. We use
differences in the signal function width to evaluate the systematic
error.

\begin{figure}
\centering
 \epsfxsize 17.5pc
 \epsfbox{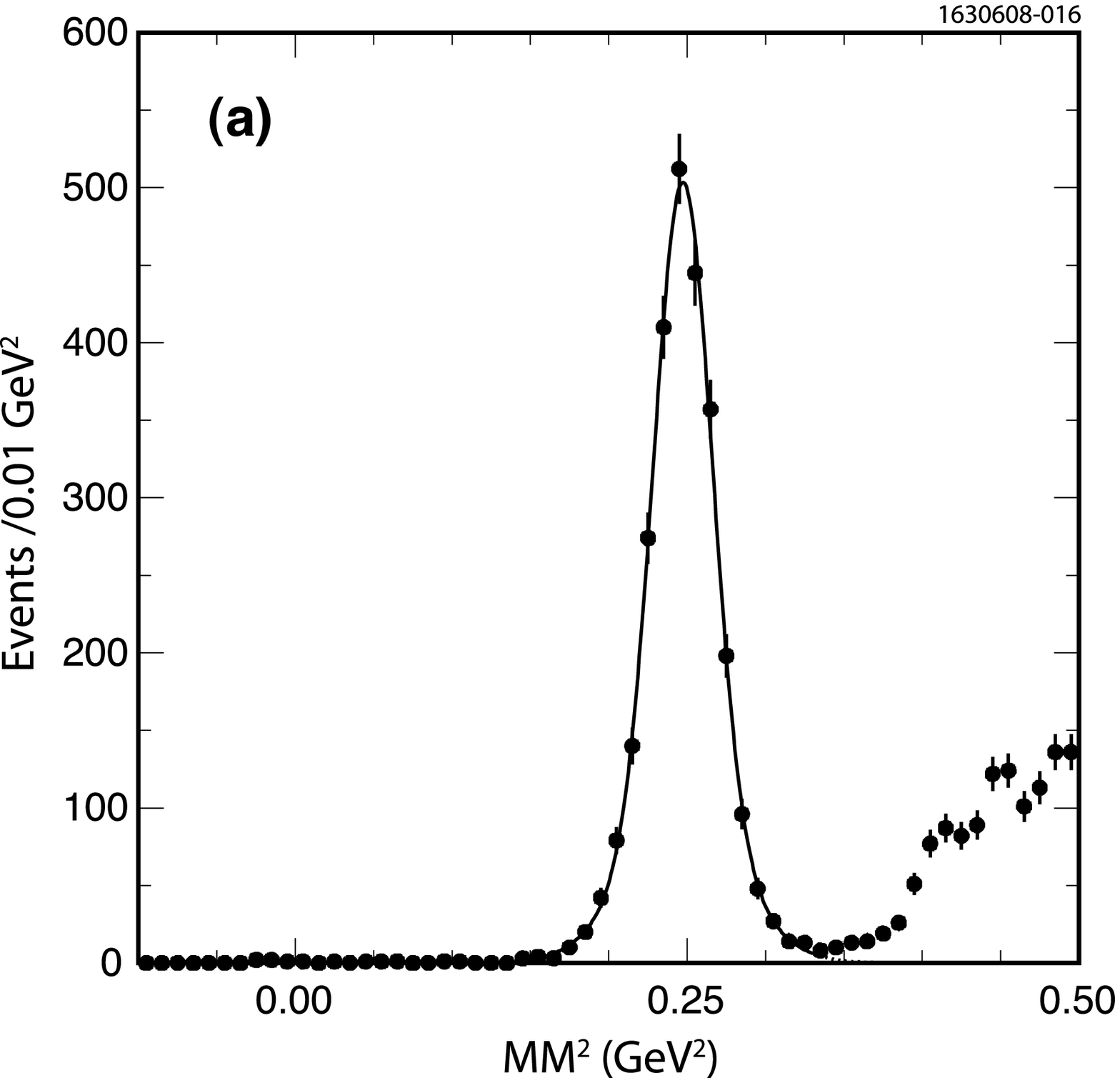}
  \epsfxsize 18pc         %
 \epsfbox{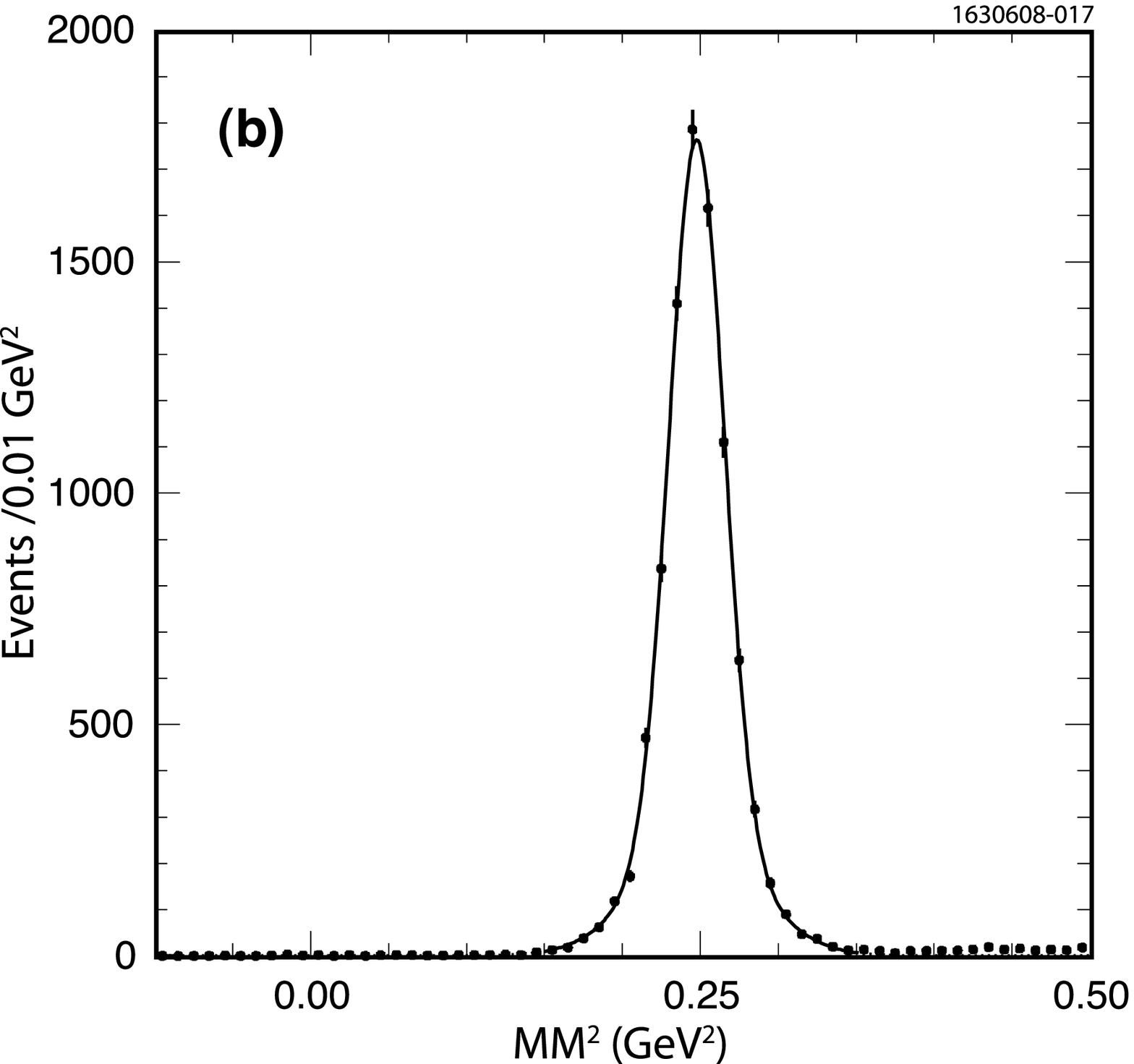}
\caption{MM$^2$ distribution for the decay $D^+\to K_S\pi^+$ from data (a),
and signal Monte-Carlo simulation (b).} \label{mc-data-check}
\end{figure}

We check our simulations by using the $D^+\to K_S\pi^+$ decay. Here
 we choose events with the same requirements as used to search for
 $\mu^+\nu$ but require one additional found $K_S$. The MM$^2$
 distribution for this final state is shown in
 Fig.~\ref{mc-data-check}(a)
 and peaks as expected at the $K_S$ mass-squared of 0.25 GeV$^2$. The
 resolution depends slightly on the fitting range, which must be specified
 since the data have a high MM$^2$ background. In the interval
$0.05<$MM$^2<0.35$ GeV$^2$, the data show a resolution of
$\sigma=0.0247\pm 0.0012$ GeV$^2$, while the Monte Carlo fit gives a
consistent value of $\sigma=0.0235\pm 0.0007$ GeV$^2$.

\begin{figure}[htb]
%\vskip 0.00cm \centerline{ \epsfxsize=3.0in
\centerline{ \epsfxsize=5.0in \epsffile{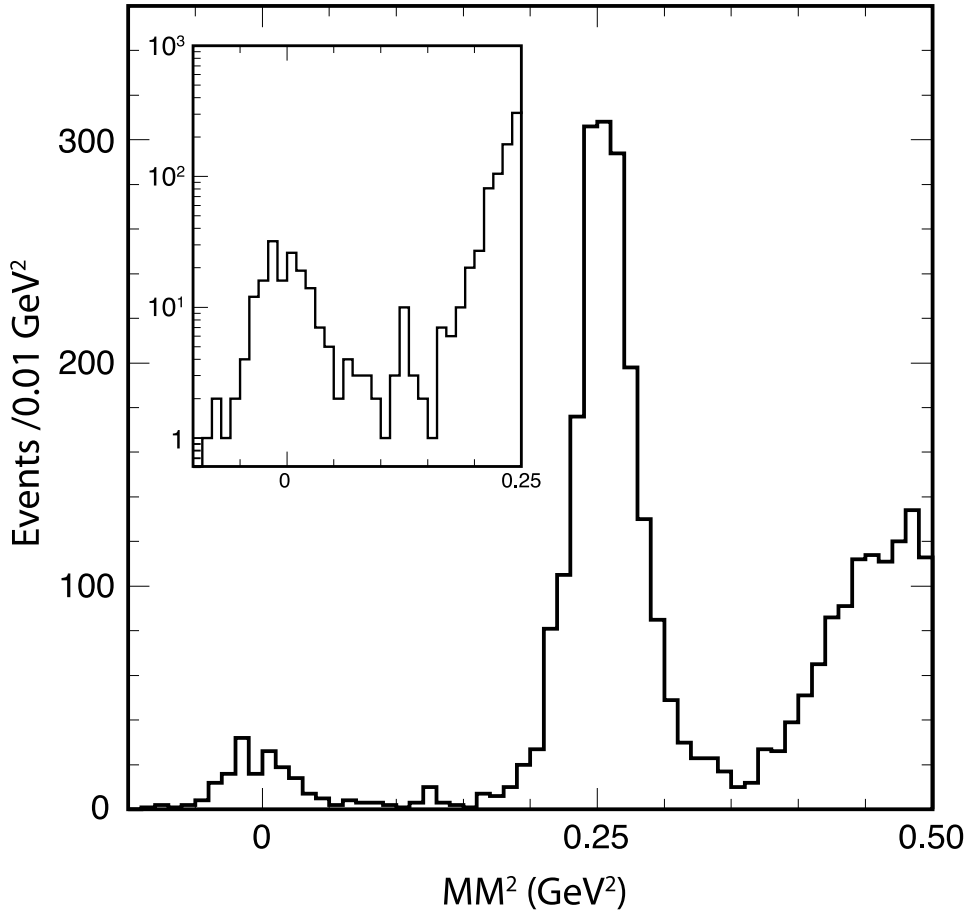} }
\caption{MM$^2$ using $D^-$ tags and one additional opposite sign
charged track and no extra energetic showers (see text). The insert
shows the signal region for $D^+\to\mu^+\nu$ on a log scale. }
\label{mm2}
\end{figure}

The MM$^2$ distributions for our tagged events requiring no extra
charged tracks besides the muon candidate and no extra showers above 250 MeV
as described above are shown in Fig.~\ref{mm2}.  We see a peak near
zero mostly due to the $D^+\to\mu^+\nu$ mode we are seeking.
The large peak centered near 0.25 GeV$^2$, far from our signal region,
results from the decay $D^+\to\overline{K}^0\pi^+$, and is
expected since many $K_L$ escape our detector.

\section{Background Considerations}
In this section we will estimate backgrounds from specific sources
and also specify shapes of several distinct background
distributions. Our procedure will be to fit the signal MM$^2$
distribution with the sum of the signal and background shapes and
then subtract off any residual backgrounds, which we will show are
very small. The signal shapes include both the $\mu^+\nu$ and
$\tau^+\nu$, $\tau^+\to\pi^+\nu$ distributions, separately.

There are several background sources we need to evaluate. These
include background from other $D^+$ modes, background from
misidentified $D^0\overline{D}^0$ events and continuum background
including that from $e^+e^-\to\gamma\psi(2S)$, termed ``radiative
return." Hadronic sources need to be considered because the
requirement of the muon depositing less than 300 MeV in the
calorimeter, while 98.8\% efficient for muons, rejects only 45\% of
pions.

We include a calculated background from $D^+\to\pi^+\pi^0$ in the
fit, both the shape and the normalization. This mode is the most
difficult to reject because the MM$^2$ peaks very close to zero, at
0.018 GeV$^2$, well within our resolution of 0.0266 GeV$^2$. It is
possible for the photons from the $\pi^0$ decay to inadvertently be
matched to the tracks from the tagging $D^-$ or be missed, even
though  at least one photon from the $\pi^+\pi^0$ mode exceeds our
250 MeV calorimeter energy requirement and should in most cases
cause such a decay to be vetoed. Both the shape in MM$^2$ and the
rate are accurately determined \cite{CLEOpipi}. Using Monte Carlo
simulation, we find efficiencies of 1.53\% and 1.06\%, for the
calorimeter energy deposition cases (i) and (ii), respectively.
(Recall case (i) is for energies less than 300 MeV, and case (ii)
for larger energy depositions.) Multiplying this efficiency by the
number of tags and branching ratio, (1.3$\pm$0.2)\%, gives a 9.2
event background. The uncertainty in the branching ratio is included
in the systematic error.

\begin{figure}[htb]
%\vskip 0.00cm \centerline{ \epsfxsize=6.0in
\centerline{\epsfxsize=4.5in \epsffile{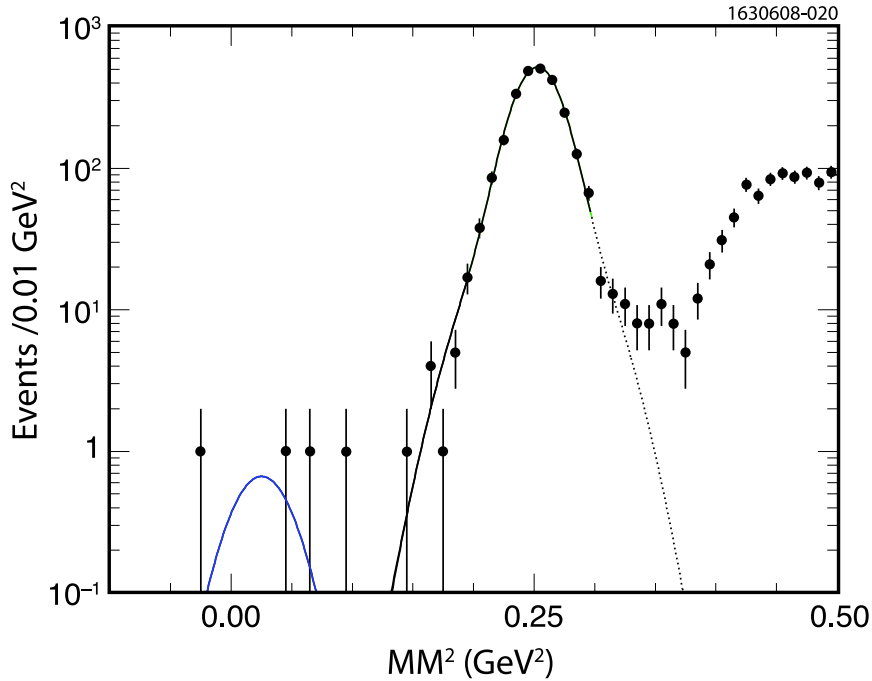}}
\caption{The MM$^2$ from events with $D^0\to K^-\pi^+$ tag and the
other $D$ decaying into two tracks, most likely $\overline{D}^0\to
K^+\pi^-$, where the kaon is ignored. The kaon peak is fit to a
double Gaussian distribution, containing 2,547 events.  The other
curve shows the expected shape for $\pi^+\pi^-$.} \label{K0tail}
\end{figure}

The $\overline{K}^0\pi^+$ mode gives a large peak in the MM$^2$
spectrum near 0.25 GeV$^2$. While it is many standard deviations from our
signal region, we need to know the shape of the tail of this
distribution. We also need to see if there are any ``pathological"
events due to non-Gaussian effects.  We use the double tag $D^0$
events where both $D$'s decay into $K^{\mp}\pi^{\pm}$ to evaluate
both effects. Here we gather a sample of single tag $K^-\pi^+$ decays using
strict $\Delta E$ and $m_{\rm BC}$ criteria, and look for events with only
two oppositely charged tracks where the ring imaging Cherenkov system (RICH)
identifies one as a $K^+$ and other as a $\pi^-$. The kaon is required
to be in the RICH solid angle but the pion can be anywhere within
$|\cos(\theta)|<0.9$, and then we ignore the kaon. The MM$^2$
distribution is shown in Fig.~\ref{K0tail}.

The fit gives us a rather good description of the shape of the
$\overline{K}^0\pi^+$ peak, especially on the low MM$^2$ side, where
the $K^0\pi^+\pi^0$ background is absent. There are 2,547 $K\pi$
events. The small numbers
 of residual events peaking
near the pion mass squared could be due to $\pi^+\pi^-$ events where
the RICH was fooled. The fake rate in the RICH has been well
measured as $(1.2\pm 0.4)$\% for pions faking kaons in the momentum
region of interest (see Appendix B). The relative branching is
${\cal B}(D^0\to\pi^+\pi^-)/{\cal B}(D^0\to K^+\pi^-)$ 3.59\% \cite{PDG}.
Thus we expect 1.1 $\pi^+\pi^-$ events.
There are three observed events consistent with being in the signal
region near MM$^2$ of zero GeV$^2$. These three events then can be
either background associated with $K^+\pi^-$ events or $\pi^+\pi^-$
events. Our best estimate is that 1.9 of them are background. By
normalizing the background based on the number of $\overline{K}^0\pi^+$
events in the MM$^2$ spectrum, we expect 1.3 events as the background
from this non-Gaussian effect in case (i) events.

\begin{figure}[htb]
%\vskip 0.00cm \centerline{ \epsfxsize=6.0in
\centerline{ \epsfxsize=4.0in \epsffile{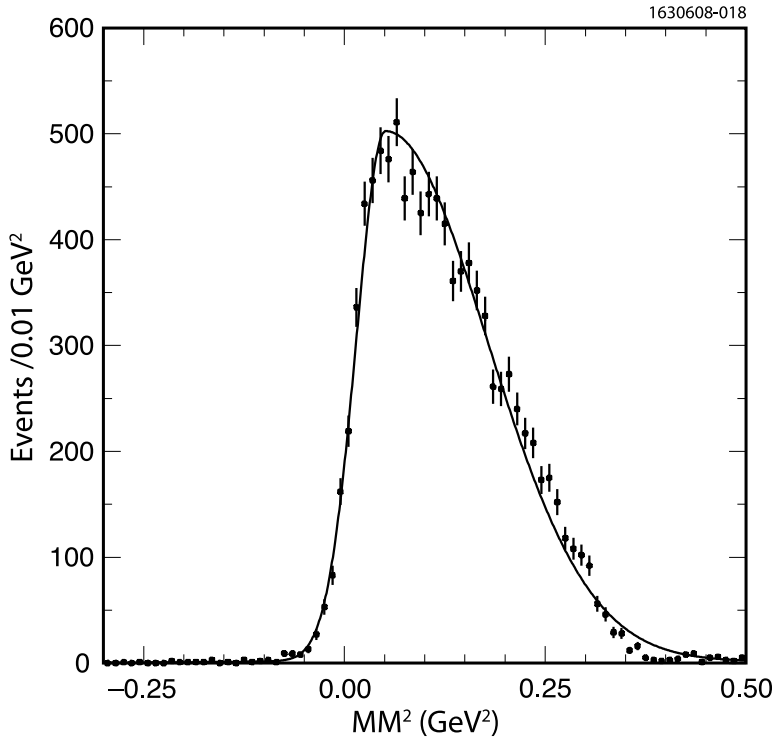}
} \caption{The simulated MM$^2$ from $D^+\to\tau^+\nu$, $\tau^+\to
\pi^+\nu$ fit to the sum of two Gaussian distributions, whose widths
are allowed to vary on both sides of their respective maximums.}
\label{mm2-case1-tautopinu-op}
\end{figure}

The only significant non-$\mu^+\nu$ population in the signal region
arises from $D^+\to\tau^+\nu$. Out of 10,000 simulated events with
$D^-$ tags, we find events in the $\mu^+\nu$ signal region only when
$\tau^+\to \pi^+\nu$. Because of the small $D^+$-$\tau^+$ mass
difference, the $\tau^+$ is almost at rest in the laboratory frame
and thus the $\pi^+$ has relatively large momentum causing the
MM$^2$ distribution to populate preferentially the low MM$^2$ region, even
though there are two missing neutrinos in this case.  Thus, we
generate a shape from Monte Carlo specifically for this one decay
sequence as shown in Fig.~\ref{mm2-case1-tautopinu-op}.

%In this analysis we fit the MM$^2$ distribution including the
%$\overline{K}^0\pi^+$ peak.

Other backgrounds from $\tau^+$ decays include additional missing
particles. We form a shape consisting of a sum of the $\tau^+$ decay
modes $\rho^+\bar{\nu}$, $\mu^+\nu\bar{\nu}$ and other ``similar"
modes $\rho^+\pi^0$, and $\pi^0\mu^+\nu$. All the relevant branching
ratios are known, where we take the $D^+\to\tau^+\nu$ rate by
multiplying our previous $\mu^+\nu$ result by 2.65, the Standard Model
prediction. We use this
shape to describe these backgrounds; we do not, however, fix the
normalization in the fit.

We have also checked the possibility of other $D^+D^-$ decay modes
producing background  with an equivalent 1.7 fb$^{-1}$ Monte Carlo
sample; we find no additional events. The {$D^0\overline{D}^0$} and
continuum backgrounds are also evaluated by analyzing Monte Carlo
samples corresponding to 4.1 and 3.0 fb$^{-1}$, respectively. To
normalize our Monte Carlo events to our data sample we used
$\sigma_{D^0\overline{D}^0}=3.7$ nb \cite{sighad} and $\sigma_{\rm
continuum}=18$ nb. We also found no events in our analysis of a
simulated radiative return sample equivalent to 2.7 fb$^{-1}$. Our
total additional background is 2.4$\pm$1.0 events, with the
individual components listed in Table~\ref{tab:Dpback}.
\begin{table}[htb]
\begin{center}
\caption{Backgrounds from additional sources, not contained in the
fitting functions. } \label{tab:Dpback}
\begin{tabular}{lc}\hline\hline
    Mode  & \# of Events \\\hline
Continuum  & 0.8$\pm$0.4\\
$\overline{K}^0\pi^+$  &1.3$\pm$0.9\\
$D^0$ modes & 0.3$\pm$0.3\\
\hline
Sum &2.4$\pm$1.0\\
\hline\hline
\end{tabular}
\end{center}
\end{table}

\section{Branching Ratio and Decay Constant}

\begin{figure}[htp]
%\vskip 0.00cm \centerline{ \epsfxsize=6.0in
\centerline{\epsfxsize=5.0in\epsffile{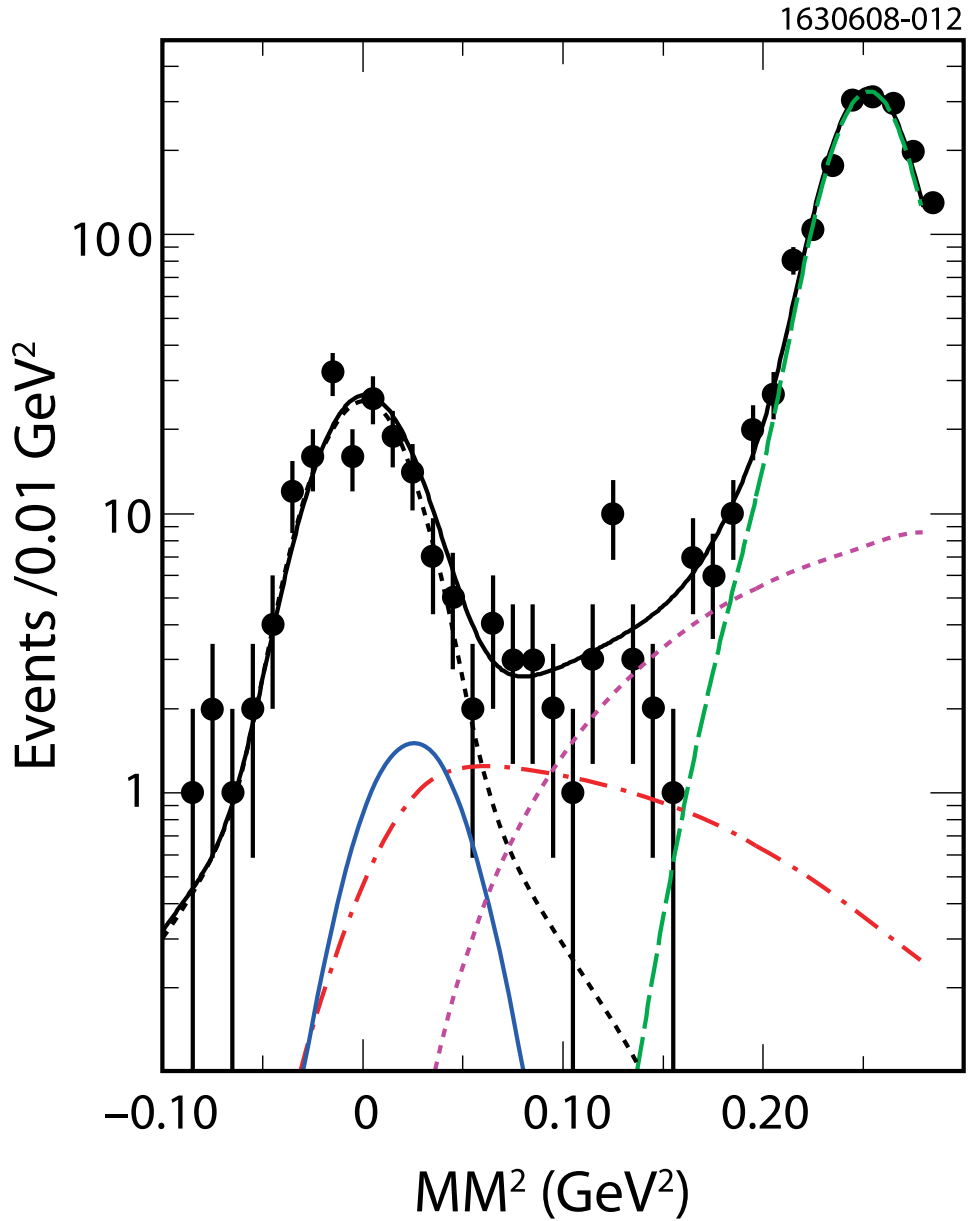}}
 \caption{Fit to the MM$^2$ for case (i). Here the ratio of
 $\tau^+\nu$, $\tau^+\to\pi^+\bar{\nu}$ to $\mu^+\nu$ events is fixed
 to the SM value. The points with error bars
 show the data. The black (dashed) curve centered at zero
 shows the signal $\mu^+\nu$ events. The dot-dashed (red) curve that peaks around
 0.05 GeV$^2$ shows the $D^+\to\tau^+\nu$, $\tau^+\to\pi^+\bar{\nu}$
 component.  The solid (blue) Gaussian shaped
 curve centered on the pion-mass squared shows the residual
 $\pi^+\pi^0$ component. The dashed (purple) curve that falls to
 zero around 0.03 GeV$^2$ is the
 sum of all the other background components, except the $\overline{K}^0\pi^+$
 tail which is shown by the long-dashed (green) curve that peaks up at
 0.25 GeV$^2$. The solid (black) curve is the sum of all the
 components.}
 \label{case1-taunufix}
\end{figure}

We preform a binned maximum liklihood fit to the case (i) MM$^2$ distribution up to a MM$^2$ of 0.28
GeV$^2$. Beyond that value other final states such as $\eta\pi^+$
and $K^0\pi^+\pi^0$ begin to contribute.  The fit shown in
Fig.~\ref{case1-taunufix} contains separate shapes for signal,
$\pi^+\pi^0$, $\overline{K}^0\pi^+$, $\tau^+\nu$ ($\tau^+\to
\pi^+\bar{\nu})$, and the background cocktail described above. Here
we assume the Standard Model ratio of 2.65 for the ratio of the
$\tau^+\nu/\mu^+\nu$ component and constrain the area ratio of these
components to the product of 2.65 with ${\cal{B}}(\tau^+\to
\pi^+\bar{\nu}$)=(10.90$\pm$0.07)\% \cite{PDG} and the 55\% probability that
the pion deposits $<$300 MeV in the calorimeter.
 The normalization of the $\pi^+\pi^0$
component is also fixed at 9.2 events, the product of the number of
tags, times the branching fraction, times the 1.53\% detection
efficiency. The normalization of the additional background shape
described above is allowed to float.

\afterpage{\clearpage}

\begin{figure}[htb]
%\vskip 0.00cm \centerline{ \epsfxsize=6.0in
\centerline{\epsfxsize=5.0in\epsffile{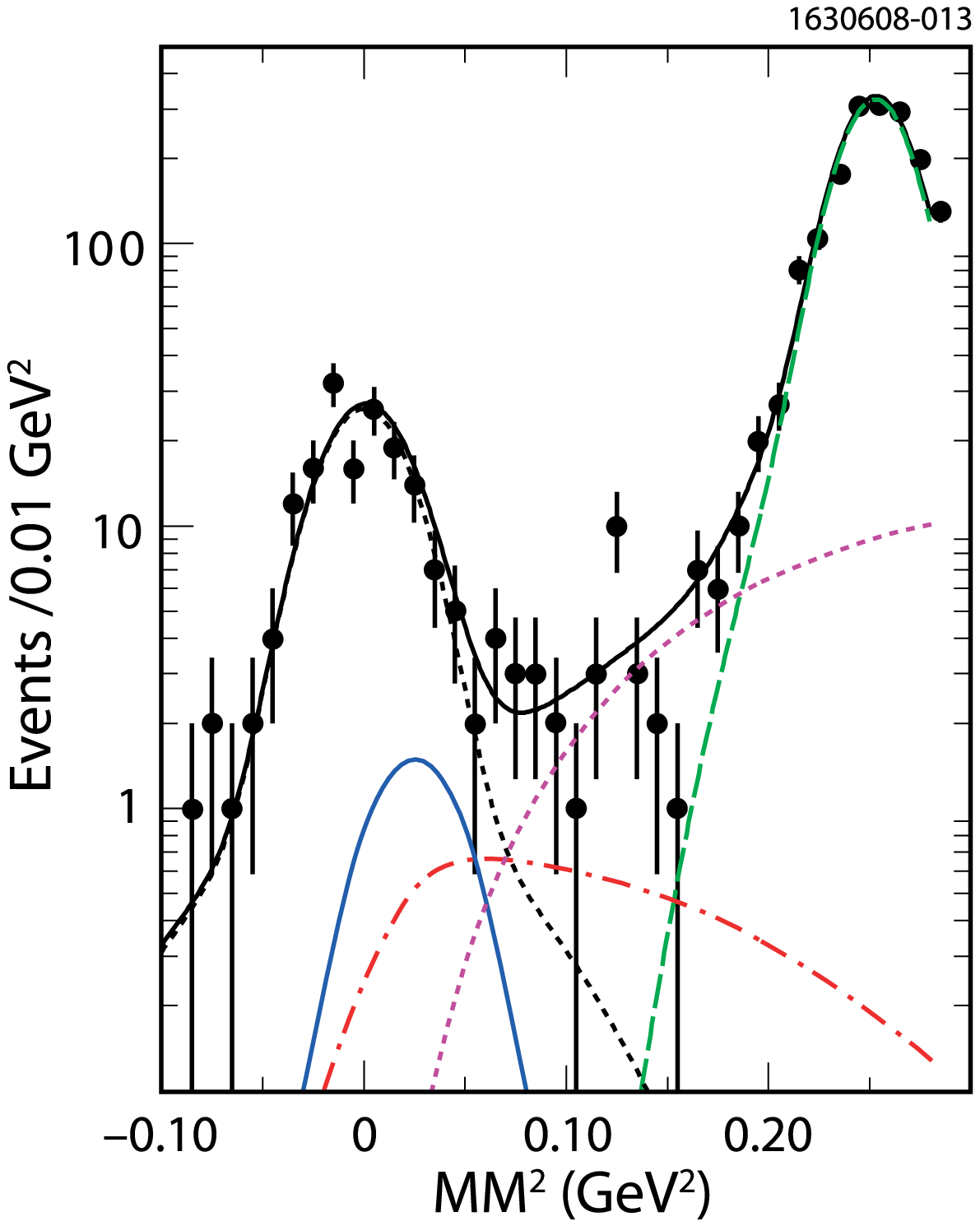}}
 \caption{Fit to the MM$^2$ for case (i) allowing the $\tau^+\nu$,
 $\tau^+\to\pi^+\bar{\nu}$ component to float. The points with error bars
 show the data. The black (dashed) curve centered at zero
 shows the signal $\mu^+\nu$ events. The dot-dashed (red) curve that peaks around
 0.05 GeV$^2$ shows the $D^+\to\tau^+\nu$, $\tau^+\to\pi^+\bar{\nu}$
 component.  The solid (blue) Gaussian shaped
 curve centered on the pion-mass squared shows the residual
 $\pi^+\pi^0$ component. The dashed (purple) curve that falls to
 zero around 0.03 GeV$^2$ is the
 sum of all the other background components, except the $\overline{K}^0\pi^+$
 tail which is shown by the long-dashed (green) curve that peaks up at
 0.25 GeV$^2$. The solid (black) curve is the sum of all the
 components.}
 \label{case1-floatbgd-op}
\end{figure}

The fit yields 149.7$\pm$12.0 $\mu^+\nu$ signal events and 25.8
$\tau^+\nu$, $\tau^+\to\pi^+\bar{\nu}$ events (for the entire MM$^2$ range). We can also perform
the fit allowing the $\tau^+\nu$, $\tau^+\to\pi^+\bar{\nu}$
component to float. (See Fig.~\ref{case1-floatbgd-op}.) Then we find
153.9$\pm$13.5 $\mu^+\nu$ events and 13.5$\pm$15.3 $\tau^+\nu$,
$\tau^+\to\pi^+\bar{\nu}$ events, compared with the 25.8 we expect
in the Standard Model. Performing the fit in this manner gives a
result that is independent of the SM expectation of the
$D^+\to\tau^+\nu$ rate. To extract a branching fraction, in either
case, we subtract off the 2.4$\pm$1.0 events determined above to be
additional backgrounds, not taken into account by the fit, and
divide by the product of the efficiency and the number of tags.

\afterpage{\clearpage}

The detection efficiency of 81.8\% for the single muon includes the
tracking and particle identification efficiencies, the probability of the
crystal energy being less than 300 MeV, and the 95.9\% efficiency of
not having another unmatched shower in the event with energy greater
than 250 MeV; the latter is determined from the data presented
in Table~\ref{tab:double} of Appendix A. The systematic errors on the branching
ratio are listed in Table~\ref{tab:eff}.
\begin{table}[htb]
\begin{center}
\caption{Systematic errors on the $D^+ \to \mu^+ \nu$ branching
ratio.} \label{tab:eff}
\begin{tabular}{lc}\hline\hline
     &Systematic errors (\%) \\ \hline
%MC statistics &0.2  \\
Track finding &0.7 \\
PID cut &1.0 \\
MM$^2$ width & 0.2\\
Minimum ionization cut &1.0 \\
Number of tags& 0.6\\
Extra showers cut & 0.4  \\
Radiative corrections & 1.0\\
 Background & 0.7\\\hline
Total &2.2\\
 \hline\hline
\end{tabular}
\end{center}
\end{table}

The systematic error on the MM$^2$ fit is determined by changing the
signal shape and the fitting range. The difference in signal shapes
between the $K_S\pi^+$ data and Monte Carlo is 0.0012$\pm$0.0014
GeV$^2$. We refit the case (i) data while increasing $\sigma$ by
0.0012~GeV$^2$, 0.0024~GeV$^2$, and finally letting $\sigma$ float.
(We fix the $\tau^+\nu/\mu^+\nu$ ratio.) The resulting numbers of
events change from our baseline by +0.41, +0.79 and +0.26 events,
respectively. This allows us to set the 0.2\% systematic error from
this source.

The track finding and
particle identification efficiencies associated with the single
muon are determined by comparing
selected samples formed using partial reconstruction \cite{abs}
to the Monte Carlo simulation. We
include the particle identification because we do veto identified
kaons as muon candidates.

\begin{figure}[htb]
%\vskip 0.00cm \centerline{ \epsfxsize=6.0in
\centerline{\epsfxsize=5.0in\epsffile{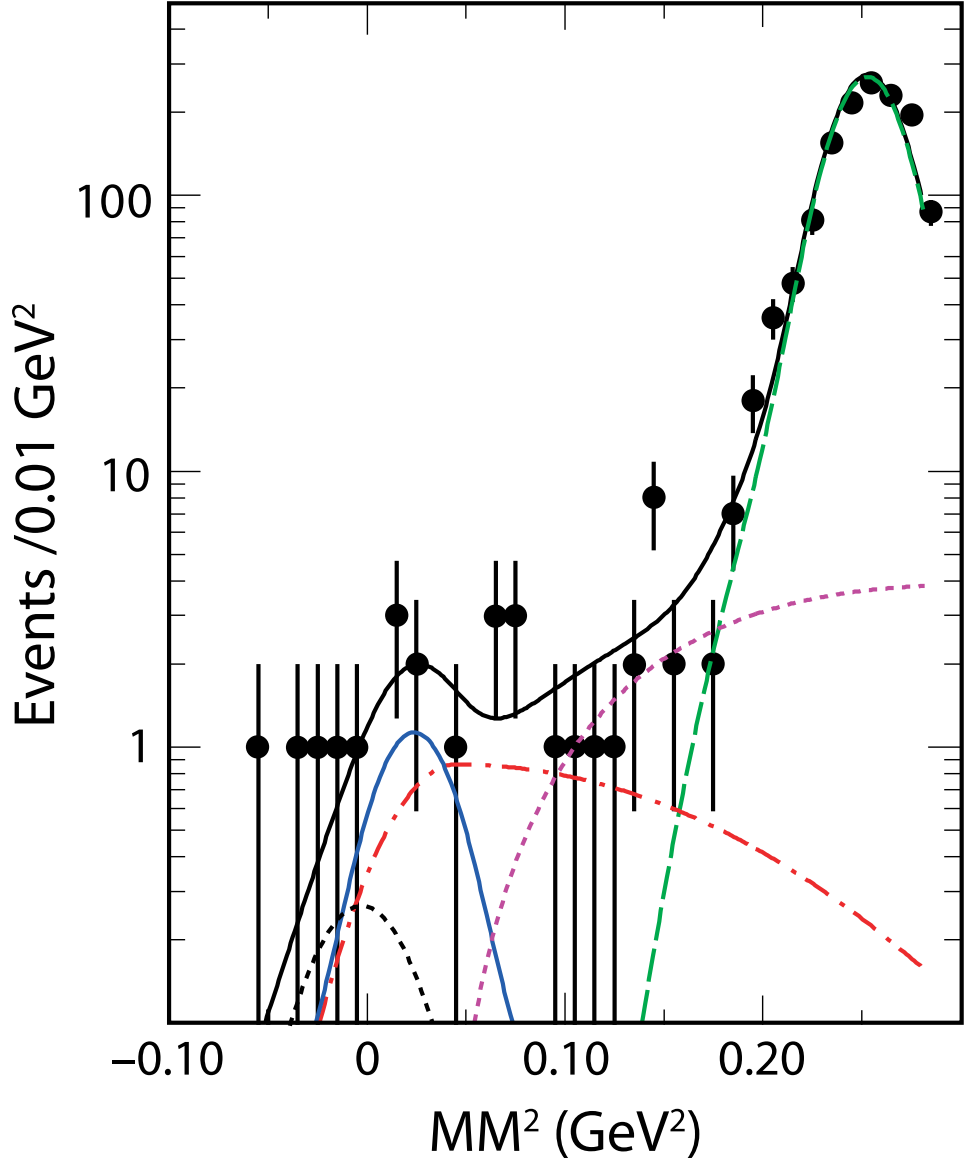}}
\caption{Fit to the MM$^2$ for case (ii) which has little $\mu^+\nu$ signal
contribution, and tests our understanding of the background.  The points with error bars
 show the data. The $\mu^+\nu$ component, the black (dashed) curve (almost invisibly small)
 shows the signal $\mu^+\nu$ events fixed from the case (i) fit. The points with error bars
 show the data.
 %The black (dashed) curve centered at zero shows the signal $\mu^+\nu$ events.
 The dot-dashed (red) curve that peaks around
 0.05 GeV$^2$ shows the $D^+\to\tau^+\nu$, $\tau^+\to\pi^+\bar{\nu}$
 component.  The solid (blue) Gaussian shaped
 curve centered on the pion-mass squared shows the residual
 $\pi^+\pi^0$ component. The dashed (purple) curve that falls to
 zero around 0.03 GeV$^2$ is the
 sum of all the other background components, except the $\overline{K}^0\pi^+$
 tail which is shown by the long-dashed (green) curve that peaks up at
 0.25 GeV$^2$. The solid (black) curve is the sum of all the
 components.}
\label{case2-taunufix-op}
\end{figure}

A check of the background is provided by considering case (ii),
where more than 300 MeV is deposited in the calorimeter by the muon
candidate track. Only 1.2\% of muons pass such a requirement.
 We fit this sample as in case (i), but here fixing both
the $\mu^+\nu$ and the $\tau^+\nu$ contributions from the case (i) fit (with
the ratio of the two fixed). The normalizations of the $K^0\pi^+$ tail and the
background shape are allowed to float.
 The fit is shown in Fig.~\ref{case2-taunufix-op}.
The number of events in the signal region, MM$^2\le$0.05 GeV$^2$, is
1.7, fixed from the $\mu^+\nu$ sample, 5.4 fixed from the
$\pi^+\pi^0$, and 4.0 from the $\tau^+\nu$. This sums to 11.1
events, while we count 11 events in this region. Thus we have an
excess of -0.1$\pm$3.3 events, which is consistent with our other background
estimate of 2.4$\pm$1.0 events and gives us confidence in using this
estimate.

The branching fraction determined from fixing the
$\tau^+\nu$ contribution relative to the $\mu^+\nu$, is
\begin{equation}
{\cal{B}}(D^+\to\mu^+\nu)=(3.82\pm 0.32\pm 0.09)\times 10^{-4}~.
\end{equation}
The decay constant $f_{D^+}$ is then obtained from
Eq.~(\ref{eq:equ_rate}) using 1040$\pm$7 fs as the $D^+$ lifetime \cite{PDG} and
0.2256 as $|V_{cd}|$ \cite{Antonelli}. Our final result is
\begin{equation}
f_{D^+}=(205.8\pm 8.5\pm 2.5)~{\rm MeV}~.
\end{equation}

A somewhat less precise value is obtained by floating the $\tau^+\nu$ to
$\mu^+\nu$ ratio. That fit gives
\begin{equation}
{\cal{B}}(D^+\to\mu^+\nu)=(3.93\pm 0.35\pm 0.09)\times 10^{-4}~.
\end{equation}
The corresponding value of the decay constant is
\begin{equation}
f_{D^+}=(207.6\pm 9.3 \pm 2.5)~{\rm MeV}~.
\end{equation}
The former value is the most precise measurement in the context of
the Standard Model, while the latter does not use any Standard Model assumptions.
In both cases the additional systematic errors due to the $D^+$
lifetime measurement and the error on
$|V_{cd}|=|V_{us}|$ are negligible.

The data have already been corrected for final state radiation of
the muon, as our Monte Carlo simulation incorporates this effect
\cite{photos}.  There is however, another process where the
$D^+\to\gamma D^{*+}\to\gamma\mu^+\nu$, where the $D^{*+}$ is a
virtual vector or axial-vector meson. The $D^{*+}\to\mu^+\nu$
transition is not helicity-suppressed, so the factor $\alpha$ for
radiation is compensated by a relative factor $(M_{D^+}/m_{\mu})^2$.
Using Eq. (12) of Burdman \etal ~\cite{Burdman} and imposing the
250~MeV photon cut, we find that the radiative rate is approximately
1\%, to which we assign a $\pm$1\% systematic error. This is
essentially the same calculation done by Dobrescu and Kronfeld for
$D_s^+\to\mu^+\nu$ decays \cite{Dobrescu-Kron}. (The results shown
above for the branching fractions and $f_{D^+}$ are all radiatively
corrected; the branching fractions have been reduced by 1\%.)

\section{Search for {\boldmath $D^+\to \tau^+\nu$}}
\begin{figure}[htb]
%\vskip 0.00cm \centerline{ \epsfxsize=6.0in
\centerline{\epsfxsize=5.0in\epsffile{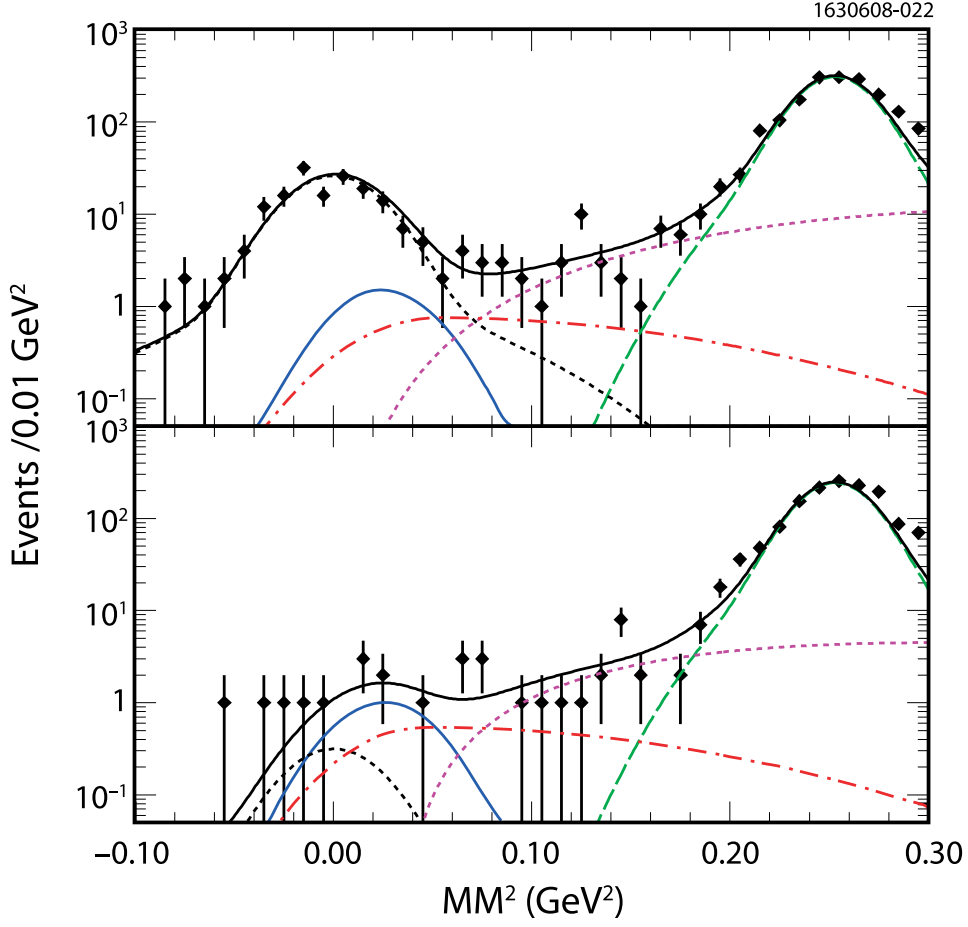}}
\caption{Fit to the MM$^2$ for cases (i) (top) and (ii) (bottom) with the $\tau^+\nu$
components fixed in the ratio 55/45.  The points with error bars
 show the data. The black (dashed) curve centered at zero
 shows the signal $\mu^+\nu$ events. The dot-dashed (red) curve that peaks around
 0.05 GeV$^2$ shows the $D^+\to\tau^+\nu$, $\tau^+\to\pi^+\bar{\nu}$
 component.  The solid (blue) Gaussian shaped
 curve centered on the pion-mass squared shows the residual
 $\pi^+\pi^0$ component. The dashed (purple) curve that falls to
 zero around 0.03 GeV$^2$ is the
 sum of all the other background components, except the $\overline{K}^0\pi^+$
 tail which is shown by the long-dashed (green) curve that peaks up at
 0.25 GeV$^2$. The solid (black) curve is the sum of all the
 components.}
\label{two-plots}
\end{figure}

We also use our data to perform a search for the $\tau^+\nu$ final state.
Here we do a simultaneous binned maximum liklihood fit to both the case (i) and case (ii) data fixing
the ratio of the $\tau^+\nu$ final state to be 55/45 in the two cases, determined by the
relative acceptances for the 300 MeV calorimeter energy requirement.
The fits are shown in Fig.~\ref{two-plots}.

The fit yields a sum of 27.8$\pm$16.4 $\tau^+\nu$, $\tau^+\to\pi^+\overline{\nu}$
events for the entire MM$^2$ range. To be conservative in setting an upper limit,
we assume all events are signal and do not subtract additional backgrounds
from this yield. We include
the small systematic errors from the fitting procedure in our calculations.
We find

\begin{equation}
{\cal{B}}(D^+\to\tau^+\nu)< 1.2\times 10^{-3}
\end{equation}
at 90\% confidence level, and
the ratio to the $\mu^+\nu$ rate divided by the Standard Model
expectation of 2.65 is
\begin{equation}
{\Gamma(D^+\to\tau^+\nu)\over {2.65\cdot\Gamma(D^+\to\mu^+\nu)}}< 1.2
\end{equation}
also at 90\% confidence level.

\section{Search for {\boldmath $D^+$} Decay into an Positron plus Neutrino}

We use the same tag sample. Candidate positrons are selected on the
basis of a likelihood ratio constructed from three inputs: the ratio
between the energy deposited in the calorimeter and the momentum
measured in the tracking system, the specific ionization $dE/dx$
measured in the drift chamber, and RICH information.
 Other criteria remain the same, except that
we require that the positron candidate track be in the calorimeter
barrel with $|\cos\theta|<0.81$. We do not find any candidates
allowing us to set a limit

\begin{equation}
{\cal{B}}(D^+\to e^+\nu)<8.8\times 10^{-6}~{\rm at~90\%~c.l.},
\end{equation}
which is three orders of magnitude above the SM prediction.

\section{Conclusions}
The result shown here represents the only precision measurement of
the pseudoscalar decay constant $f_{D^+}$.  We have significantly
improved our previous results. The statistical error has been
reduced by almost a factor of two. The systematic errors remain
small. This result uses all the CLEO-c data collected at the $\psi(3770)$ and,
as such, all previous results are superseded.

 The branching fraction, assuming the Standard Model ratio
 for $\tau^+\nu/\mu^+\nu$ is
\begin{equation}
{\cal{B}}(D^+\to\mu^+\nu)=(3.82\pm 0.32\pm 0.09)\times 10^{-4}~,
\end{equation}
and the decay constant is
\begin{equation}
f_{D^+}=(205.8\pm 8.5\pm 2.5)~{\rm MeV}~.
\end{equation}

If, on the other hand, we allow the $\tau^+\nu$ contribution to
float, we find
\begin{equation}
{\cal{B}}(D^+\to\mu^+\nu)=(3.93\pm 0.35\pm 0.09)\times 10^{-4}~.
\end{equation}
The corresponding value of the decay constant is
\begin{equation}
f_{D^+}=(207.6\pm 9.3 \pm 2.5)~{\rm MeV}~.
\end{equation}
These results are all radiatively corrected.

 Our new values are
consistent with our previous measurement \cite{CLEODptomunu}, as
well as the upper limit set by Mark III \cite{MarkIII}, and the
results based on reported yields of 1 and 2.7 events from BES I and
II \cite{Rong2}, respectively. We also determine
$f_{D_s}/f_{D^+}=1.326\pm 0.075$, using the world average value of
absolute measurements for $D^+_s\to\ell^+\nu$ as compiled by Rosner
and Stone \cite{Rosner-Stone}, where we include the radiative
correction also on the $D_s^+$ rate.

 Our
result for $f_{D^+}$, is consistent with the most accurate
unquenched lattice QCD (LQCD) calculation of Follana \etal ~who give
a value of (207$\pm$4) MeV. This implies that the somewhat greater
than three standard deviation discrepancy of the experimental
measurements of $f_{D_s}=(273\pm 10)$ MeV \cite{Rosner-Stone} with
the Follana \etal ~prediction of (241$\pm$3) MeV cannot be explained
by how they handle the charm quark in their calculation. In fact,
since the $s$ quark is heavier than the $d$ quark, it should be
easier for lattice calculations to predict $f_{D_s}$ than $f_{D^+}$
\cite{private}. It may be the case that physics beyond the Standard
Model is raising the value of $f_{D_s}$ in one of the ways suggested
by Dobrescu and Kronfeld \cite{Dobrescu-Kron}, or via R-parity
violating supersymmetry \cite{Kundu-Nandi}. Other Standard Model
based predictions are listed in Table~\ref{tab:Models}.

\begin{table}[htb]
\begin{center}
\caption{Theoretical predictions of $f_{D^+}$ and
$f_{D_s^+}/f_{D^+}$} \label{tab:Models}
\begin{tabular}{lcl}\hline\hline
    Theory  &  $f_{D^+}$ (MeV)          &  $f_{D_s^+}/f_{D^+}$           \\\hline
LQCD (HPQCD+UKQCD) \cite{Lat:Follana} &
$207\pm 4$&$1.164\pm 0.011$ \\
LQCD (Fermilab+MILC) \cite{Lat:Milc} &
$201\pm 3 \pm 17 $&$1.24\pm 0.02\pm 0.07$ \\

QL (QCDSF) \cite{QCDSF}& $206\pm 6\pm 3\pm 22$&
   $1.07\pm 0.02\pm 0.02$\\
 QL (Taiwan) \cite{Lat:Taiwan} & $235\pm 8\pm
14$&
   $1.13\pm 0.03\pm 0.05$\\
 QL (UKQCD) \cite{Lat:UKQCD} & $210\pm
10^{+17}_{-16}$&
   $1.13\pm 0.02^{+0.04}_{-0.02}$\\
 QL \cite{Lat:Damir} & $211\pm
14^{+2}_{-12}$&
   $1.10\pm 0.02$\\
 QCD Sum Rules \cite{Bordes} & $177\pm 21$& $1.16\pm
0.01\pm 0.03$\\
 QCD Sum Rules \cite{Chiral} & $203\pm
20$& $1.15\pm 0.04$\\
Field Correlators \cite{Field} & $210\pm 10$& $1.24\pm 0.03$\\
QCD Sum Rules \cite{Sumrules} & $195\pm 20$ & \\
Relativistic Quark Model \cite{Quarkmodel} & 234& 1.15 \\
Potential Model \cite{Equations} & 238  & 1.01 \\
Isospin Mass Splittings \cite{Isospin} & $262\pm 29$ & \\
\hline\hline
\end{tabular}
\end{center}
\end{table}

It is possible in some models of new physics that there is a
difference in the $\mu\nu$ decay rate between $D^+$ and $D^-$
mesons, due to a CP violating interaction \cite{Zwicky}. Separating
our data into these two classes we find 228,945$\pm$551 $D^+$ tags
and 231,107$\pm$552 $D^-$ tags. Fitting the data by fixing the
relative $\tau^{\pm}\nu$, $\tau^{\pm}\to\pi^{\pm}\nu$ contribution
relative to $\mu^{\pm}\nu$, we find 76.0$\pm$8.6 $\mu^+\nu$ events
and 64.8$\pm$8.1 $\mu^-\nu$ events. The resulting CP violating
asymmetry is
\begin{equation}
A_{\rm CP}\equiv
\frac{\Gamma(D^+\to\mu^+\nu)-\Gamma(D^-\to\mu^-\bar{\nu})}
{\Gamma(D^+\to\mu^+\nu)+\Gamma(D^-\to\mu^-\bar{\nu})} =
0.08\pm0.08~.
\end{equation}
At 90\% confidence level the limits are $-0.05 <A_{\rm CP} <0.21$.

We do not find positive evidence of the decay $D^+\to\tau^+\nu$. Our
limit is
\begin{equation}
{\cal{B}}(D^+\to\tau^+\nu)< 1.2\times 10^{-3}
\end{equation}
at 90\% confidence level, and
the ratio to the $\mu^+\nu$ rate, divided by the Standard Model
expectation of 2.65 is
\begin{equation}
{\Gamma(D^+\to\tau^+\nu)\over {2.65\cdot\Gamma(D^+\to\mu^+\nu)}}< 1.2
\end{equation}
also at 90\% confidence level.

 Some
non-standard models predict significant rates for the helicity
suppressed decay $D^+\to e^+\nu$ \cite{Denu}. Our upper limit of
$8.8\times 10^{-6}$ at 90\% c.l. restricts these models.

\section{Acknowledgments}

We gratefully acknowledge the effort of the CESR staff in providing
us with excellent luminosity and running conditions.
D.~Cronin-Hennessy and A.~Ryd thank the A.P.~Sloan Foundation. This
work was supported by the National Science Foundation, the U.S.
Department of Energy, the Natural Sciences and Engineering Research
Council of Canada, and the U.K. Science and Technology Facilities
Council. We thank C. Davies, A. Kronfeld, P. Lepage,  P. Mackenzie,
R. Van de Water, and Roman Zwicky for useful discussions.

\afterpage{\clearpage}
\newpage

\newpage
\appendix
\section{Determination of the Efficiency of the 250
MeV Criteria on Additional Photons}

Although we do not expect more than a few percent inefficiency due
to rejecting events with an additional neutral energy cluster
$>$ 250 MeV, we do not want to incur a large systematic error due to
this potential source. Therefore we perform a full five-constraint
kinematic fit to the double tag event samples, where one $D$ decays
into $K^{\mp}\pi^{\pm}\pi^{\pm}$ and the other into one of the other
tag modes. The constraints are that the total energy sum to twice
the beam energy, the total three momentum be zero, and the
invariant masses of the two $D$ candidates be equal. We do not
require them to equal the known $D^+$ mass. The result of this fit
is a common $D$ candidate mass and a $\chi^2$. Restricting our
samples to low $\chi^2$ virtually eliminates all backgrounds at the
expense of some signal. Specifically, we require that the
probability of $\chi^2$, for five constraints be greater than 1\%,
which eliminates 32\% of all event candidates.
 The numbers of events
in the decay modes we use are listed in Table~\ref{tab:double}.

\begin{table}[htb]
\begin{center}
\caption{Numbers of $D^+ D^-$ events and the efficiency for the
first mode when an extra photon $>$ 250 MeV is also required.}
\label{tab:double}
\begin{tabular}{llccc}\hline\hline
    Mode 1 &  Mode 2           &  Events & $N_{\rm lost}~(E_{\gamma > 250~{\rm MeV}})$
    & $\epsilon_{250}$(\%) of Mode 1\\\hline
$K^+\pi^-\pi^-$  & $K^-\pi^+\pi^+$ &4389&431&95.0$\pm$0.2\\
$K^+\pi^-\pi^-\pi^0$  & $K^-\pi^+\pi^+$ &2590&208&96.8$\pm$0.6\\
$K_S\pi^- $  & $K^-\pi^+\pi^+$ &1255&112&95.9$\pm$0.8\\
$K_S\pi^-\pi^-\pi^+ $  & $K^-\pi^+\pi^+$ &1885&153&96.8$\pm$0.7\\
$K_S\pi^-\pi^0$  & $K^-\pi^+\pi^+$ &2648&205&97.7$\pm$0.5\\
$K^+K^-\pi^-$  & $K^-\pi^+\pi^+$ &714&75& 94.2$\pm$1.1\\\hline
\multicolumn{2}{l}{Weighted Average}  & & & 95.9$\pm$0.2\\

\hline\hline
\end{tabular}
\end{center}
\end{table}

To first order the fully reconstructed $D^+D^-\to (K^+\pi^-\pi^-$)
 ($K^-\pi^+\pi^+$) can be considered the superposition of two single
 tag $D^+\to\mu^+\nu$ candidate events where the single tag is
 $K^+\pi^-\pi^-$.
 Then the efficiency of the 250 MeV cut in the $K\pi\pi$ mode is
 given by
 \begin{equation}
\epsilon_{250}^{K\pi\pi}=\sqrt{(1-N_{\rm lost}/N_{K\pi\pi-K\pi\pi})}.
\end{equation}

 We then combine the large and precise
$K^-\pi^+\pi^+$ mode with each of the other tags in turn, where
\begin{equation}
\epsilon_{250}^{\rm mode}=(1- N_{\rm
lost}/N_{K\pi\pi-{\rm mode}})/\epsilon_{250}^{K\pi\pi}.
\end{equation}
 This
method ensures that the number of interactions of particles with
material is the same as in the tag sample
used for the $\mu^+\nu$ analysis.

The results are listed in Table~\ref{tab:double}. The numbers of
events listed are those with a $\chi^2$ cut applied.  The overall
efficiency for accepting the double tag event requiring that there
not be any photons above 250 MeV is given along with the derived
efficiency for each mode. The weighted average over all of our tag
modes is (95.9$\pm$0.2$\pm$0.4)\%. The
systematic error arises only from the consideration that we have
analyzed a situation corresponding to two overlapping tags rather
than one tag plus a muon.

\section{RICH Particle Identification Efficiencies}
For two-body decays of $D$ mesons, most of the particle identification
ability in CLEO comes from the RICH detector. Information on the angle of detected Cherenkov
photons is translated into a likelihood of a given photon being due
to a particular particle. Contributions from all photons associated
with a particular track are then summed to form an overall
likelihood denoted as ${\cal L}_i$ for each particle hypothesis. To
differentiate between pion and kaon candidates, we use the
difference: $-2\log({\cal L_{\pi}})+2\log({\cal L}_K$). A value
of zero is used to distinguish between the two possibilities. We require a minimum of three
Cherenkov photons.

Here we use a selected sample of $D^0\overline{D}^0\to
K^{\mp}\pi^{\pm}+K^{\pm}\pi^{\mp}$ decays. We use $-0.0194<\Delta
E<0.0175$ GeV and $1.8617<m_{\rm BC}<1.8673$ GeV for both candidates.
This is essentially a background free sample. We expect only
$K^-\pi^+K^+\pi^-$ decays since doubly Cabibbo suppressed decays are
forbidden due to quantum correlations and the mixing rate as
measured is small enough not to allow us to see any events. The
momentum distribution of the tracks is flat between 700 MeV/c and 1
GeV/c.

The results are shown in Table~\ref{Kpi-Kpi}. The first column labeled ``No ID"
gives the number of $K^-\pi^+;K^+\pi^-$ pairs called right sign (RS) and the number of
$K^-\pi^+;K^-\pi^+$ (or $K^+\pi^-;K^+\pi^-$) pairs that are wrong sign (WS) using only
the kinematical constraints of $\Delta E$ and $m_{\rm BC}$ given above. The subsequent
columns show the results of applying the RICH particle identification criterion to
identify only the kaons, only the pions and then both kaons and pions.

\begin{table}[htb]
\begin{center}
\caption{Results of RICH identification on double tag events. RS
indicates right sign and WS indicates wrong sign events.}
\label{Kpi-Kpi}
\begin{tabular}{lcccccccc}\hline\hline
\multicolumn{1}{l}{Mode}&\multicolumn{2}{c}{No ID} &
\multicolumn{2}{c}{Single $K$} & \multicolumn{2}{c}{Single $\pi$} &
\multicolumn{2}{c}{Double ID} \\
&(RS) & (WS)  &(RS) & (WS) &  (RS) & (WS) &  (RS) &   (WS)
\\\hline
$K^-\pi^+$;  $K^{\pm}\pi^{\mp}$ &1896&914&1717&11&1846&24
 & 1673&  1\\
\hline\hline
\end{tabular}
\end{center}
\end{table}

 The relevant results are summarized as:
\begin{itemize}
\item The pion efficiency is (97.3$\pm$0.3)\%.
\item The kaon efficiency is (90.6$\pm$0.7)\%.
\item The rate of pions faking kaons is $(1.2\pm0.4^{+0}_{-0.1})$\%.
\item The rate of kaons faking pions is $(2.6\pm0.5^{+0}_{-0.1})$\%.
\end{itemize}

The one doubly identified wrong sign event could be a mixed event,
although that is rather unlikely. We use it to assign a negative
systematic error on the fake rates in case there is background in
our sample.


\begin{thebibliography}{99}
\bibitem{abs}
See Ref.~\cite{sighad}.
The actual values used here are larger because the momentum spectrum
is different.
\end{thebibliography}
\end{document}